# Formal Semantic Annotations for Models Interoperability in a PLM environment

Y. Liao*. M. Lezoche*. H. Panetto*. N. Boudjlida**. E. Rocha Loures***

*{* Université de Lorraine, CRAN, UMR 7039, CNRS, Boulevard des Aiguillettes B.P.70239,
54506 Vandoeuvre-lès-Nancy, France; (e-mail: yongxin.liao, mario.lezoche, herve.panetto@univ-lorraine.fr).
** Université de Lorraine, CRAN, UMR 7039, CNRS, Boulevard des Aiguillettes B.P.70239,
54506 Vandoeuvre-lès-Nancy, France; (e-mail: nacer.boudjlida@loria.fr)
*** Pontifical Catholic University of Paraná, Brezil; (e-mail: eduardo.loures@pucpr.br)}*

**Abstract:** Nowadays, the need for system interoperability in or across enterprises has become more and more ubiquitous. Lots of research works have been carried out in the information exchange, transformation, discovery and reuse. One of the main challenges in these researches is to overcome the semantic heterogeneity between enterprise applications along the lifecycle of a product. As a possible solution to assist the semantic interoperability, semantic annotation has gained more and more attentions and is widely used in different domains. In this paper, based on the investigation of the context and the related works, we identify some existing drawbacks and propose a formal semantic annotation approach to support the semantics enrichment of models in a PLM environment.

## 1. INTRODUCTION

In manufacturing enterprises, the Product Lifecycle Management (PLM) approach has been considered as an essential solution for improving the product competitive ability. It aims at providing a shared platform that brings together different enterprise systems at each stage of a Product Life Cycle (PLC) in or across enterprises (Ameri et al., 2005). Although the main software companies are making efforts to create tools for offering a complete and integrated set of systems, most of them have not implemented all of the systems. Finally, they do not provide a coherent integration of the entire information system. This results in a kind of "tower of Babel" managed by many stakeholders in an enterprise, or even in a network of enterprises. The different peculiarities of those stakeholders, who operate on those systems, are then over increasing the issue of interoperability.

The objective of this paper is to deal with the interoperability problems, mainly the issue of semantic interoperability, by proposing a formal semantic annotation method to support the mutual understanding of the semantics of the shared and exchanged models in a PLM environment. The remainder of this paper is organized as follows: Section 2 presents an overview of the research context and discusses the issue of semantic interoperability; Section 3 surveys the related works that made use of semantic annotations to deal with the interoperability issues and identifies the existing drawbacks among those researches. Section 4 presents the semantic annotation formalization proposals, suggestion and verification mechanisms, and a semantic annotation framework; Section 5 presents a case study to demonstrate the applicability and the use of the proposed solution; Section 6 concludes this paper and highlights future research directions.

## 2. RESEARCH CONTEXT

The concept of the Product Life Cycle (PLC) has been introduced since the 1950s (Rink et al.,1979), and it is a biological metaphor that describes every phase a product goes through, from the first initial requirement until it is retired and disposed of. In the meantime, along with the advent and the evolution of Computer Aided Design (CAD) systems, the problems of locating the required data and losing control of change process associated with these data have gradually appeared. As a solution, Product Data Management (PDM) systems have been developed and introduced for supporting easy, quick and secured access to valid data during the product design phase (Ameri et al., 2005). However, as it is pointed out in (Elgueder et al, 2010), the data produced by CAD systems do not cover all the information that is related to the whole product life cycle (from the requirement specification to dismantling information). The PLM solution, proposed during the 1990s, provides support to the processes of capturing, representing, retrieving and reusing both engineering and non-engineering aspects of knowledge along the entire product life cycle. It intends to facilitate the knowledge management in or across enterprises (Ameri et al., 2005). Therefore, the knowledge concerning the product life cycle, which we named PLC-related knowledge, has become one of the critical concepts in a PLM solution.

Knowledge is an awareness of things that brings to its owner the capability of grasping the meanings (semantics) from the information (Ackoff, 1989). In this work, knowledge is considered as a kind of intangible thing, which has to be made perceptible and afterward to be expressed under various kinds of representations. Knowledge representation is the result of embodying the knowledge from its owner's mind into some explicit forms. We consider that all the relevant resources produced by different stakeholders through the variety of enterprise systems are all knowledge representations, such as requirement documents, product design models, control interface designs, process models, data models, observation videos and so on. Therefore, in a PLM environment, these multifaceted forms of knowledge representations act as the carriers of PLC-related knowledge and as the basis for collaboration activities along the product life cycle.

Interoperability serves as a foundational role to support collaboration. In the compilation of IEEE standard computer glossaries (IEEE, 1991), the interoperability is defined as "The ability of two or more systems or components to exchange information and to use the information that has been exchanged". Therefore, the systems need to unambiguously interpret the exchanged information (Boudjlida et al, 2008). (Euzenat, 2001) categorized five possible levels of interoperability: encoding level, lexical level, syntactic level, semantic level and semiotic level. Semantic interoperability is the ability to ensure that the exchanged information has got the same meaning considering the point of view of both the sender and the receiver (Pokraev et al., 2007). In the context of PLM, stakeholders with different background have to work together on the exchanged knowledge representations and take decisions based on them. In order to cope with this issue there are two important obstacles that need to be overcome: (1) The implicit semantics that is necessary for understanding a knowledge representation is not be made explicit; (2) The lack of semantics mechanisms to verify the correctness of explicit semantics in the exchanged knowledge representation.

The Ontology knowledge formalization (Gruber, 1993), which is a kind of common agreement on the conceptualization of terms in a specific domain of interest, is usually considered as a possible solution to deal with these two obstacles (Boudjlida et al, 2001). Being a way to realize the semantic enrichment, the application of semantic annotation not only use the formal and shared knowledge that is represented in ontologies to make explicit implicit semantics, but also give the possibility to perform the semantics verification for those knowledge representations that are not initially designed with this ability. In this paper, there are two important aspects of the semantics that are made explicit by a semantic annotation: (1) The domain semantics, which describes the context and the meaning of an annotated element in a specific domain; (2) The structure semantics, which describes the interrelations between the annotated element and the other elements that related to it in a knowledge representation.

Before we proceed to the identification of problems and the proposition of some solutions we need to declare three hypotheses:

(H1) All the knowledge that is needed for the semantic enrichment of models has already been captured, represented and formalized into ontologies.

(H2) The corresponding interconnections among all the used ontologies have already been prepared through certain methods.

(H3) The semantic similarity between two objects can be compared through certain methods.

The support for these hypotheses can be provided by related researches in the corresponding domains. The research community, which are working on knowledge discovery (Polanyi, 1966), conversion (Nonaka, 1994), and formalization (Gruber, 1993), can give support to the hypotheses H1. Taking advantages from the researches about ontology matching (Maedche, 2002), mapping (Doan, 2003), and merging (Stumme, 2001), hypotheses H2 is possible to be achieved. A number of researchers, such as (Patil, 2004), have been committed themselves in the evaluation of semantic similarities. Based on these hypotheses, we focus our research work on proposing a solution to formalize the semantic annotation for the semantic enrichment of models in a PLM environment.

## 3. RELATED WORKS

In the face of various needs, different literatures have been proposed to use different ontologies to annotate various kinds of model in diverse ways. Enterprise modelling is a process that tries to capture and represent knowledge from different aspects of a system of interest and for activating the interoperations in or across enterprises. In this research work, we focus our inquiry on a PLM environment where all different types of models along the product life cycle are considered as the targets of semantic enrichments. These models are always created with particular perspectives and expressed in a given modelling notation (or description language). The interoperations among those systems not only require that models can be exchanged and operated on, but also demand an unambiguous understanding of the exchanged models.

From the representation point of view, a model "is often presented as a combination of drawings and text. The text may be in a modeling language or in a natural language" (Miller, 2003). The mutual understanding of a model requires not only the understanding of the semantics of "combination of drawing" (structure semantics) but also the semantics of the "text" (domain semantics). Therefore, the ontologies employed by the semantic enrichment need to capture and represent both aspects of knowledge. In this research work, two aspects of ontologies are categorised and can be used to support the semantic enrichment of models in a PLM environment: PLC-related ontologies and Meta-model ontologies. PLC-related ontologies represent the PLC-related knowledge. For example, to mention only a few, SCOR-Full ontology (Zdravković et al., 2011), and MSDL Ontology (Ameri, 2011). Meta-model ontologies represent the model constructs knowledge. They can be used to express the structure semantics of annotated elements in a model that are related to the interrelations between their counterpart components in its meta-model. Such as Petri net Ontology (Gašević, 2006), BPMN Ontology (Ghidini, 2008) and so on. Because the development of ontologies is not our research focus the MSDL ontology and BPMN ontology are employed to support the semantic enrichment in the validation of our proposition.

With the supports of the ontologies, semantic annotations could be widely used in many contexts. (Uren et al., 2006) reviewed and classified the existing semantic annotation systems as four kinds: manual annotation, automatic annotation, integrated annotation environments, and On-demand annotation. Task Group 4 of the INTEROP project (Boudjlida, 2006) proposed a general schema the semantic annotation of all enterprise models to enable both semantic-based and model-based interoperability between collaborating actors. Through the main purpose of Semantic Annotation for WSDL and XML Schema (SAWSDL) (W3C, 2005) is for the annotation of Web services, it can also be used to annotate the models that are stored in the format of XML. We found that despite lots of efforts have been made in semantic annotation researches, a number of existing drawbacks still need to be noted:

(1) The formalization of semantic annotations is not the focus in research (Bergamaschi et al., 2011), where it is only considered as a kind of "is a" association between an annotated object and an ontology concept. Meanwhile, some specific semantic annotation models are proposed by research (Attene, 2009), (Li et al., 2012) and (Di Francescomarino, 2011). However, these models are difficult to be reused in other researches but the studied ones.

(2) Making explicit the domain semantics is the only concern in research (Bergamaschi, et al., 2011) and (Li et al., 2012), where the structure semantics is ignored. The advantages of making explicit the structure semantics have been acquired by (Boudjlida, 2006), that used it to express modelling construct and support models transformations. In (Attene et al., 2009), it is used to support the automatic computation of relations between features in the model. However, among all these usages, the structure semantics and domain semantics are defined and used separately. There is a lack of research that combines both semantics together in the inference process.

(3) The verification of the correctness of those semantic annotations normally needs human involved. The research (Di Francescomarino, 2011) is the only one that proposed a mechanism to assist this verification process. However, it only verifies the types of the annotated elements but not the semantics they contain.

After all, based on the investigation of related works, a number of requirements for our proposed solution can be identified: (1) It should provide a general semantic annotation structure model that is able be used to formalize semantic annotations for different kinds of models; (2) It should discover the possibility of using both structure and domain semantics together in the inference process; (3) It should provide some mechanisms to assist the detection of the inconsistencies between semantic annotations and the identification of the conflicts between annotated elements; (4) It should provide a way to guide annotators in how to apply the formal semantic annotations and how to benefit from those semantic annotations; (5) It should provide a framework to support the semantic enrichment of models along the product life cycle. In the next section, the proposed solution that follows these requirements is presented.

## 4. PROPOSED SOLUTION

In order to address existing drawbacks and meet the listed requirements, in this section, we propose a formal approach to assist the semantic enrichment of models in a PLM environment. The essential elements of a semantic annotation are not clearly identified in current semantic annotation researches. To better formalize semantic annotations, we first present a meta-model of the semantic annotation, then we will present two kinds of semantic blocks that can be used to support the formal definition of the semantic annotation and the creation of reasoning rules. At the end we will propose the semantic annotation formal definitions.

The knowledge understanding represented by a model needs, not only the domain semantics, embedded in the model contents, but also the structure semantics, embedded in the modelling constructs. In order to define the meta-model of the semantic annotation, several important concepts that are used throughout this section need to be reviewed.

Models in a PLM environment are always expressed in some kinds of modelling languages with designer's specific peculiarities. This results in the implicit, or possibly ambiguously explicit, semantics that is not easily intelligible by the humans or the machines. In this research work, all kinds of models throughout a product life cycle are considered as Target Knowledge Representations (TKRs) for the semantic enrichment. Ontology represents a real-world semantics that enables human to use meaningful terminologies as machine processable contents.

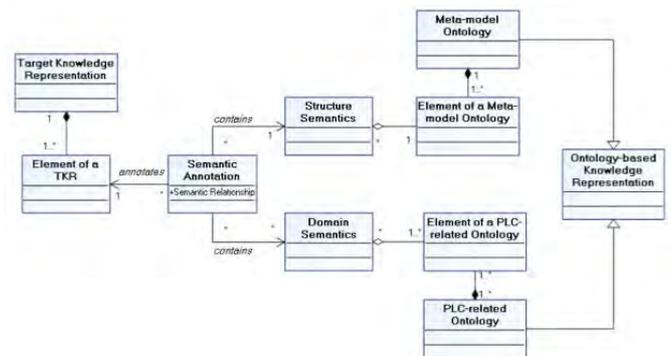

Figure 1. The Meta-model of the Semantic Annotation

In this research work, two kinds of ontologies (PLC-related and Meta-model ontologies) are considered as Ontology-based Knowledge Representations (OKRs) to support the semantic enrichment of models. The Semantic Annotation is acting as a bridge to formally describe the semantic relationships between TKRs and OKRs. In this research work, two aspects of semantics (domain and structure semantics) are made explicit through the semantic enrichment.

The meta-model of the semantic annotation is presented in Figure 1, which describes the main components of a semantic annotation and their relationships.

Taking advantages from this meta-model we propose a semantic block delimitation method that will be used as a basis to support the proposition of formal definitions and the creation of reasoning rules.

The concept of "semantic block" is adopted from the research of (Yahia et al., 2012), in which, it represents a kind of aggregation of semantics. In their research, a semantic block is composed by a minimal number of mandatory concepts that are needed to express the full semantics of an appointed concept. In this work, we extend the semantic block definition to cover the relations among those selected concepts. A semantic block is considered as a shape (segment) of a model that contains a number of selected entities and corresponding relations among them. Two kinds of semantic blocks can be categorized based on their objectives:

1. Semantic Blocks for Semantics Description: the delimitation method supports the creation of a "Domain Semantics" through delimitating one or more "Element of a PLC-related Ontology" from one or more "PLC-related Ontology". The generated semantic block is used to describe the domain semantics of an "Element of a TKR" based on the semantics that it aggregates.

2. Semantic Blocks for Semantics Substitution: the delimitation method supports the creation of a substitute through delimitating one or more "Element of a TKR" from one "Target Knowledge Representation" based on the "Structure Semantics" that they express. The produced semantic block is used as a substitute of those "Element of a TKR" it aggregates and acts as a new entity or a new relation in the "Target Knowledge Representation".

Let $A$ be a set of entities in a model. Let $B \subseteq A \times A$ be a set of binary relations. Given $a_i, a_j \in A$, we say that $a_i$ is relative to $a_j$ through $b_{i,j} = (a_i, a_j) \in B$. We call $a_i$ the domain of $b_{i,j}$ and $a_j$ the range of $b_{i,j}$.

Since the relations among entities in ontologies are already explicit, the delimitation of semantic blocks can be applied directly. Let $a_{i_0} \in A$ be a selected entity, which is named as the "main concept", and $A_{a_{i_0}} \subseteq A$ be a set of selected entities that are associated to $a_{i_0}$. Let $B_{a_{i_0}} \subseteq B$ be a set of relations among those selected entities. Mathematically, $A_{a_{i_0}}$ is defined as:

$A_{a_{i_0},0} = \{a_{i_0}\}$;

$A_{a_{i_0},1} = \{a_{i_1} \in A | \exists b_{i_0,i_1} \in B_{a_{i_0}}, a_{i_0} \in A_{a_{i_0},0}, (a_{i_0}, a_{i_1}) = b_{i_0,i_1}\}$;

$A_{a_{i_0},2} = \{a_{i_2} \in A | \exists b_{i_1,i_2} \in B_{a_{i_0}}, a_{i_1} \in A_{a_{i_0},1}, (a_{i_1}, a_{i_2}) = b_{i_1,i_2}\}$;

…

$A_{a_{i_0},n} = \{a_{i_n} \in A | \exists b_{i_{n-1},i_n} \in B_{a_{i_0}}, a_{i_{n-1}} \in A_{a_{i_0},n-1}, (a_{i_{n-1}}, a_{i_n}) = b_{i_{n-1},i_n}\}$;

$A_{a_{i_0}} := \bigcup_n A_{a_{i_0},n}$

According to user define selection methods that are applied during the creation of this kind of semantic blocks, an appropriate subset of the $A$ can be determined. Then the semantic block of the entity $a_{i_0}$ is defined as a pair:

$SB_{a_{i_0}} := (A_{a_{i_0}}, B_{a_{i_0}})$,

where every entity in $A_{a_{i_0}}$ can be attained by $a_{i_0}$ through, at least, one path and all the relations in the paths are contained in $B_{a_{i_0}}$.

The (a) in the Figure 2 depicts a part of an ontology that contains explicit relations and the (b) in the Figure 2 shows the semantic block $SB_{a_{7vii}}$ of the main concept $a_{7vii}$, which can be used to describe the domain semantics of the element it annotates. Taking advantage from this kind of semantic blocks, the annotators can, with a certain degree of freedom, delimitate an appropriate semantics that they needed in the OKRs.

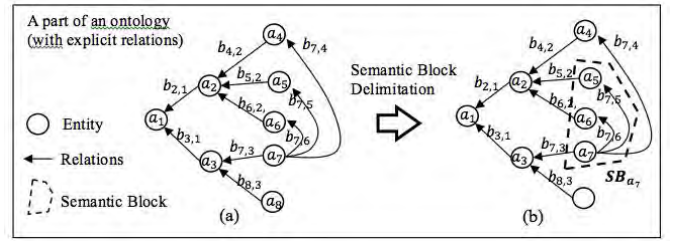

Figure 2. An Example of the Semantic Block for Semantics Description

Due to the relations among the entities are implicit in enterprise models, the delimitation cannot be applied directly. As shown in the Figure 3, two processes are proposed as follows:

(1) Relation Explicitation.

There are two general rules in the relation explicitation process:

Every model element is represented as an entity of the set $A$.

A relation $b_{i,j} = (a_i, a_j) \in B$ is created between $a_i \in A$ and $a_j \in A$, when the model element that is represented by $a_i$ is related to the model element that is represented by $a_j$.

(2) Semantic Block Delimitation

This kind of semantic blocks can be further divided into two categories depending on the role it acts: as an entity or as a relation. In this paper, we only use and introduce the latter category. The restrictions for the delimitation are generated as follows:

A semantic block that acts as a new relation between $a_i \in A$ and $a_j \in A$. Let $A_{a_i,a_j} \subseteq A$ be a set of selected entities and let $B_{a_i,a_j} \subseteq B$ be a set of relations among $a_i$, $a_j$ and the entities in $A_{a_i,a_j}$. In order to substitute the semantics of its contents, it needs to satisfy the following three conditions:

$A_{a_i,a_j}$ does not contain $a_i$ and $a_j$. That is $a_i, a_j \notin A_{a_i,a_j}$.

For every entity $a_k$ in the $A_{a_i,a_j}$, at least one entity $a_l$ exists in $A_{a_i,a_j}$ that has a relation $b_{k,l}$ in $B_{a_i,a_j}$ to $a_k$. That is

$\forall a_k \in A_{a_i,a_j}, \exists a_l \in A_{a_i,a_j}, \exists b_{k,l} \in B_{a_i,a_j}, \ s.t. \ (a_k, a_l) = b_{k,l}$.

Beside $a_i$ and $a_j$, for every binary relation $b_{k',l'}$ in the $B_{a_i,a_j}$, the entities that appear in the domain and range of $b_{k',l'}$ are the entities in the $A_{a_i,a_j}$. That is

$\forall b_{k',l'} \in B_{a_i,a_j}, \ (a_{k'}, a_{l'}) = b_{k',l'} \Rightarrow a_{k'}, a_{l'} \in A_{a_i,a_j} \cup \{a_i\} \cup \{a_j\}$

Then the semantic block $SBR_{a_i,a_j}$ is defined as a pair:

$$SBR_{a_i,a_j} := (A_{a_i,a_j}, B_{a_i,a_j})$$

The (a) in the Figure 3 shows a part of a process model that contains implicit relations. The (b) in the Figure 3 shows the represented entities and explicit relations. The (c) in the Figure 3 shows the semantic block $SBR_{a_1,a_4}$ that merges the semantics of its contents and acts as a new relation between $a_1$ and $a_4$. Taking advantages from this kind of semantic block, a combination of elements in the TKR can be delimitated and act as new entities or relations to assist the creation of reasoning rules.

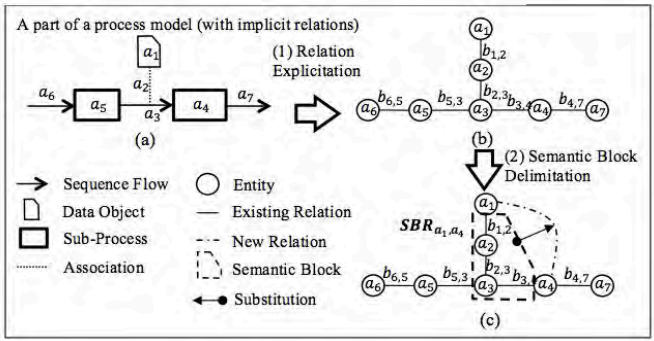

Figure 3. An Example of the Semantic Block for Semantics Substitution

Based on the formal definitions of semantic annotation that are proposed in (Liao et al., 2013) an improved version is presented in this section. Let $E$ be the set of elements in a TKR and $e_i$ be one of the elements in $E$.

*Definition 1.* An ontology is a formal and shared understanding of some domains of interest, which specifies the concepts and the relationships that can exist for an agent or a community of agents (Gruber, 1993). Let $o_x$ represent an ontology, which is formalized by a triple:

$o_x := (C_{o_x}, R_{o_x}, A_{o_x})$,

where $C_{o_x}$ is a set of concepts; $R_{o_x}$ is a set of relationships; $A_{o_x}$ is a set of axioms. Let $oall_{o_x}$ be the set that contains all the elements from the set $C_{o_x}$ and $R_{o_x}$. An ontology element $oe_{o_{x_y}}$ is represented as:

$oall_{o_x} = \{oe_{o_{x_y}} | oe_{o_{x_y}} \in C_{o_x} \cup R_{o_x}\}$.

*Definition 2.* A meta-model is a model that specifies the concepts, relationships and rules to model a model. Let $mm_x$ denote a meta-model, which is defined as a triple:

$mm_x := (C_{mm_x}, R_{mm_x}, RU_{mm_x})$,

where $C_{mm_x}$ is a set of concepts; $R_{mm_x}$ is a set of relationships; $RU_{mm_x}$ is a set of rules.

Let $mmo_x$ be an ontology that represents the meta-model $mm_x$, which is defined as:

$mmo_x := (C_{mmo_x}, R_{mmo_x}, A_{mmo_x})$.

*Definition 3.* The domain semantics of a TKR is made explicit by one or more PLC-related ontologies. Let $PO$ be the set of PLC-related ontologies and $P$ be the set of selected ontology element sets from the powerset of all ontology elements of $PO$, which is defined as:

$\bigcup_{o_x \in PO} oall_{o_x} = \{oe_{o_{x_y}} | (\exists o_x)(o_x \in PO \wedge oe_{o_{x_y}} \in oall_{o_x})\}$,

$P \subseteq \mathcal{P}(\bigcup_{o_x \in O} oall_{o_x})$.

*Definition 4.* The structure semantics of a TKR is made explicit by a meta-model ontology $mmo_x$. Let MME be the set that contains all the elements from the set $C_{mmo_x}$. An ontology element $mme_l$ is defined as:

$MME := \{mme_l | mme_l \in C_{mmo_x}\}$.

*Definition 5.* Let $A$ and $B$ be two sets, any subset of $br \subseteq A \times B$ is a binary relation from A to B. Given $a \in A$ and $b \in B$, the $br$ in the notation $a \ br \ b$ is defined as,

$br := \{(a,b) | a \text{ is in the relation } br \text{ with } b\}$.

Let $dom(br)$ represent the domain of the $br$ and $ran(br)$ represent the range of the $br$, which are defined as

$dom(br) := \{a \in A | \exists b \in B, (a,b) \in br\}$,

$ran(br) := \{b \in B | \exists a \in A, (a,b) \in br\}$.

*Definition 6.* $SR_{E,P}$ is a set of binary relations that describe the semantic relationships from $E$ to $P$. Given, $e_i \in E$ and $p_j \in P$, and let $sem(e_i)$ represent the semantics of $e_i$ and $sem(p_j)$ represent the semantics of $p_j$, five subsets of the $SR_{E,P}$ are defined as follows:

$sr_\sim := \{(e_i, p_j) | sem(e_i) \text{ and } sem(p_j) \text{ are equivalent}\}$;

$sr_\supset := \{(e_i, p_j) | sem(e_i) \text{ is more general than } sem(p_j)\}$;

$sr_\subset := \{(e_i, p_j) | sem(e_i) \text{ is less general than } sem(p_j)\}$;

$sr_\cap := \{(e_i, p_j) | e_i \text{ and } p_j \text{ have common semantics}, (e_i, p_j) \notin sr_\sim \cup sr_\supset \cup sr_\subset\}$;

$sr_\perp := \{(e_i, p_j) | e_i \text{ and } p_j \text{ have not common semantics}\}$.

*Definition 7.* $MR_{E,MME}$ is a set of binary relations that describe the semantic relations from $E$ to $MME$. Given $e_i \in E$ and $mme_l \in MME$, one subset of $MR_{E,MME}$ is defined as follow:

$mr_{io} := \{(e_i, mme_l) | e_i \text{ is an instance of } mme_l\}$.

Finally, with all above-mentioned definitions, we are now ready to formally define the semantic annotation.

*Definition 8.* Let TKR, $PO$ and $mmo_x$ be given, the semantic annotation $SA$ that is associated to them is defined by a 5-tuple:

$$SA \coloneqq (E, P, SR, MME, MR),$$

where

$E$ is a set of elements from a TKR;

$P$ is a set of selected ontology element sets from a set of PLC-related ontologies $PO$, which makes explicit the domain semantics aspect of $E$;

$MME$ is a set of ontology elements from a meta-model ontology $mmo_x$, which makes explicit the structure semantics aspect of $E$;

$SR \coloneqq SR_{E,P}$;

$MR \coloneqq MR_{E,MME}$.

These formal definitions not only can be used to construct a semantic annotation schema, but also can be used as the basis for the creation of reasoning mechanisms.

In this work, the formal semantic annotations are mainly contributing in two main aspects: for assisting the creation of models and for supporting the identification of possible mistakes. Therefore, three main stages with their corresponding mechanisms are proposed for achieving these two purposes: (1) the suggestion of semantic annotations; (2) the inconsistency detection between semantic annotations; and (3) the conflict identification between annotated objects in a model.

The essence of an inconsistency is the contradictory among two or more facts that describe one common object. With the same principle, the inconsistency detection between semantic annotations is based on the comparison of two or more semantic annotations that describe the semantics of the same "Element of a TKR". Therefore, to cope with this premise, two types of semantic annotations are classified: Initial Semantic Annotations, which are directly annotated on an "Element of a TKR" by an annotator; Inferred Semantic Annotations, which are suggested to annotate an "Element of a TKR" through an inference action that is based on its related element's semantic annotations and corresponding reasoning rules. Both "Structure Semantics" and "Domain Semantics" are contributing in the annotation suggestion stage. The "Structure Semantics" is used to make explicit the implicit relations between the annotated "Element of a TKR" and its related elements. The "Domain Semantics" is used as the basis for the annotation suggestion. Two remarks need to be pointed out: (1) only the semantic relationship $sr_\subset$ and $sr_=$ can produce suggestions; (2) the semantic blocks that are nested within each other are not taken into account.

The detection of inconsistencies can be performed on the annotated element that has two or more semantic annotations. Using the case of inconsistency detection between two semantic annotations as the basis, let $e_i$ be annotated by $sa_x$ and $sa_y$, in which, $p_x$ and $p_y$ are used to make explicit the domain semantics of $e_i$. The semantic similarity comparison results between $p_x$ and $p_y$ is defined.

*Definition 9.* $PR$ is a binary relation that describes the semantic relationships from $P$ to $P$. Given $p_x, p_y \in P$, and let $sem(p_y)$ represent the semantics of $p_x$ and $sem(p_y)$ represent the semantics of $p_y$, five subsets of $PR$ are defined as follows:

$pr_\sim \coloneqq \{(p_x, p_y) | sem(p_x) \text{ and } sem(p_y) \text{ are equivalent}\}$;

$pr_\supset \coloneqq \{(p_x, p_y) | sem(p_x) \text{ is more general than } sem(p_y)\}$;

$pr_\subset \coloneqq \{(p_x, p_y) | sem(p_x) \text{ is less general than } sem(p_y)\}$;

$pr_\cap \coloneqq \{(p_x, p_y) | p_x \text{ and } p_y \text{ have common semantics}, (p_x, p_y) \notin pr_\sim \cup pr_\supset \cup pr_\subset\}$;

$pr_\perp \coloneqq \{(p_x, p_y) | p_x \text{ and } p_y \text{ have not common semantics}\}$.

As shown in the Table 1, according to the similarity comparison between two domain semantics of a common annotated element, three types of results can be identified as follows: result (a) expresses that $sa_x$ and $sa_y$ are consistent with each other; result (b) expresses that $sa_x$ and $sa_y$ are possible consistent with each other; result (c) expresses that there is an inconsistency between $sa_x$ and $sa_y$. To be more succinct, we use the concept "Others" to replace the rest of the semantic relationships in the $PR$ besides the one or several that are shown in a grid of the table.

Table 1. The Possible Results of the Inconsistency Detection

|  | $e_i\ sr_\sim\ p_x$ | $e_i\ sr_\supset\ p_x$ | $e_i\ sr_\subset\ p_x$ | $e_i\ sr_\cap\ p_x$ | $e_i\ sr_\perp\ p_x$ |
|---|---|---|---|---|---|
| $e_i\ sr_\sim\ p_y$ | (a) $p_x\ pr_\sim\ p_y$ (c) Others | (a) $p_x\ pr_\subset\ p_y$ (c) Others | (a) $p_x\ pr_\supset\ p_y$ (c) Others | (a) $p_x\ pr_\cap\ p_y$ (c) Others | (a) $p_x\ pr_\perp\ p_y$ (c) Others |
| $e_i\ sr_\supset\ p_y$ | (a) $p_x\ pr_\supset\ p_y$ (c) Others | (b) $p_x\ pr_\sim\ p_y$ $p_x\ pr_\supset\ p_y$ $p_x\ pr_\subset\ p_y$ $p_x\ pr_\cap\ p_y$ $p_x\ pr_\perp\ p_y$ | (a) $p_x\ pr_\supset\ p_y$ (c) Others | (b) $p_x\ pr_\supset\ p_y$ $p_x\ pr_\cap\ p_y$ $p_x\ pr_\perp\ p_y$ (c) Others | (a) $p_x\ pr_\perp\ p_y$ (c) Others |
| $e_i\ sr_\subset\ p_y$ | (a) $p_x\ pr_\subset\ p_y$ (c) Others | (a) $p_x\ pr_\subset\ p_y$ (c) Others | (b) $p_x\ pr_\sim\ p_y$ $p_x\ pr_\subset\ p_y$ $p_x\ pr_\supset\ p_y$ $p_x\ pr_\cap\ p_y$ (c) Others | (b) $p_x\ pr_\subset\ p_y$ $p_x\ pr_\cap\ p_y$ (c) Others | (b) $p_x\ pr_\subset\ p_y$ $p_x\ pr_\perp\ p_y$ (c) Others |
| $e_i\ sr_\cap\ p_y$ | (a) $p_x\ pr_\cap\ p_y$ (c) Others | (b) $p_x\ pr_\subset\ p_y$ $p_x\ pr_\cap\ p_y$ $p_x\ pr_\perp\ p_y$ (c) Others | (b) $p_x\ pr_\supset\ p_y$ $p_x\ pr_\cap\ p_y$ (c) Others | (b) $p_x\ pr_\sim\ p_y$ $p_x\ pr_\supset\ p_y$ $p_x\ pr_\subset\ p_y$ $p_x\ pr_\cap\ p_y$ $p_x\ pr_\perp\ p_y$ | (b) $p_x\ pr_\subset\ p_y$ $p_x\ pr_\cap\ p_y$ $p_x\ pr_\perp\ p_y$ (c) Others |
| $e_i\ sr_\perp\ p_y$ | (a) $p_x\ pr_\perp\ p_y$ (c) Others | (a) $p_x\ pr_\perp\ p_y$ (c) Others | (b) $p_x pr_\supset p_y$ $p_x\ pr_\cap\ p_y$ $p_x\ pr_\perp\ p_y$ (c) Others | (b) $p_x\ pr_\supset\ p_y$ $p_x\ pr_\cap\ p_y$ $p_x\ pr_\perp\ p_y$ (c) Others | (b) $p_x\ pr_\sim\ p_y$ $p_x\ pr_\supset\ p_y$ $p_x\ pr_\subset\ p_y$ $p_x\ pr_\cap\ p_y$ $p_x\ pr_\perp\ p_y$ |

The inconsistency detection results not only point out the inconsistencies (or possible inconsistencies) between two (or more) semantic annotations, but also can be used to identify the possible conflicts between those annotated elements in a TKR. Using the case of conflict identification between two annotated elements in a TKR as the basis, let $e_i$, $e_j$ and $e_k$ be three elements in a TKR and there is an inconsistency between $sa_x$ and $sa_y$ that are both used to annotate $e_i$. As shown in the Table 2, the possible conflicts between two annotated elements in the TKR can be identified.

Table 2. The Possible Results of Conflict Identification

| The $sa_x$ on $e_i$ is an \ The $sa_y$ on $e_i$ is an | Initial Semantic Annotation | Inferred Semantic Annotation (It is inferred from the semantic annotation of $e_j$ ) |
|---|---|---|
| Initial Semantic Annotation | | Between $e_i$ and $e_j$, one of them is possibly wrong |
| Inferred Semantic Annotation (It is inferred from the semantic annotation of $e_k$) | Between $e_i$ and $e_k$, one of them is possibly wrong | Between $e_j$ and $e_k$ (if $e_j \neq e_k$): one of them is possibly wrong |

In order to apply the above-mentioned semantic annotation proposal in a PLM environment we propose a semantic annotation framework that capture, represent and manage the knowledge related to the system of interest through the semantic annotation. An overview of the procedure to apply semantic annotations is presented in the Figure 4.

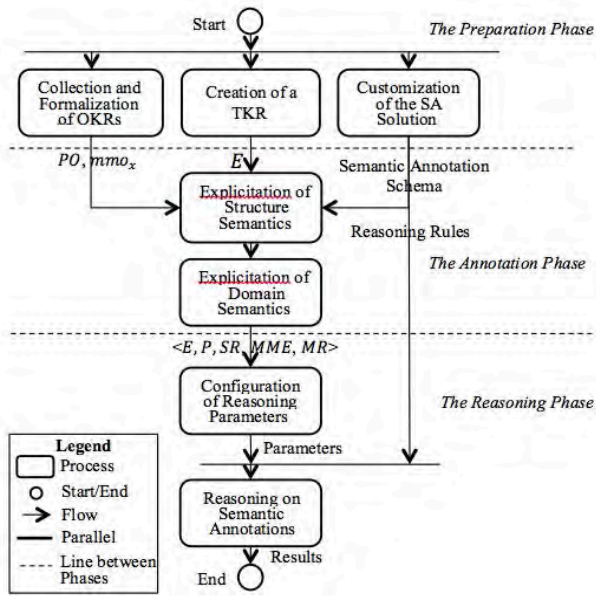

Figure 4. The General Semantic Annotation Procedure.

This workflow is divided into three phases as follows:

*The Preparation Phase.* During this phase, all the elements that are needed by both the annotation phase and the reasoning phase are prepared: (1) Creation of a TKR by a modelling system. The set of elements $E$ in this TKR are the output of this process; (2) Collection and Formalization of OKRs, in which, the ontologies are captured and formalized. The output of this process is a number of PLC-related ontologies ($PO$) and a meta-model ontology ($mmo_x$); (3) Customization of the SA Solution, in which, the formal definitions of semantic annotations and the reasoning mechanisms are used as the foundation to customize a semantic annotation schema and corresponding reasoning rules. The former one is used as a repository to conserve the semantic annotations; the letter one is used to support the inference process in the reasoning phase.

*The Annotation Phase.* During this phase, a number of semantic annotations are generated for supporting the reasoning phase: (1) Explicitation of Structure Semantics, in which, the structure semantics of a TKR, namely the interrelations between the model elements, are made explicit. (2) Explicitation of Domain Semantics, in which, the domain semantics of a TKR, namely the meaning of model contents in a domain of interest, are made explicit.

*The Reasoning Phase.* During this phase, the reasoning is performed based on the outputs of the above-mentioned two phases: (1) Configuration of Reasoning Parameters, in which, based on the semantic annotation schema and the practical situations of different TKRs, the operations that support the configuration of reasoning parameters are performed. (2) Reasoning on Semantic Annotations, in which, the reasoning is performed based on the semantic annotations, the parameters and the reasoning rules to produce inference results.

The semantic annotation procedure describes the application of the semantic enrichment solution within one single TKR. So to deal with the multiple TKRs in a PLM environment we propose a semantic annotation framework. As shown in the Figure 5, on the left side, there are a series of processes that describe a linear product life cycle, which represent the TKR Creation and Management module. On the right side, there are four main modules of this framework: the OKR Creation and Management module, the Knowledge Cloud module the Semantic Annotation and Processing Agent (SAPA) module and the Reasoning Engine module.

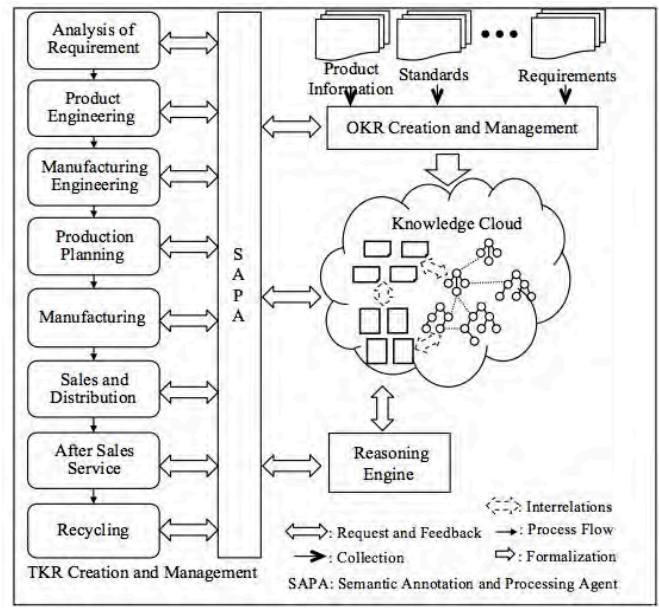

Figure 5. The Semantic Annotation Framework in a PLM Environment.

The TKR Creation and Management module is composed by a number of enterprise systems. Stakeholders in or across enterprises, during a product lifecycle use those systems to create and manage TKRs. Those systems need to provide sufficient APIs to enable the communications between themselves and the SAPA module.

The OKR Creation and Management module is in charge of capturing, formalizing and managing PLC-related knowledge and model constructs knowledge into a knowledge base, namely, Knowledge Cloud. The OKRs are supposed to be in a platform independent format, which ensures they can be imported, mapped, merged and interrelated with each other.

The Knowledge Cloud module acts as a knowledge repository. In this research work, three kinds of knowledge representations are stored in the knowledge cloud: (1) All the OKRs produced by the OKR Creation and Management module; (2) All the semantic annotations that are created by different stakeholders along the product lifecycle via the SAPA module; (3) All the reasoning rules.

The Reasoning Engine module is an external call pattern-matching search engine. It performs the inferences on the knowledge that is stored in the semantic annotations, in the OKRs and in the reasoning rules.

The Semantic Annotation and Processing Agent (SAPA) model is mainly in charge of the semantic relationships definition process. It also acts as a mediator to support the communications between various kinds of modelling systems in different processes of the PLC and the three other modules in the semantic annotation framework:

Between the Knowledge Cloud module and the modelling systems: according to the annotation requests from the stakeholders, it queries the Knowledge Cloud and then provides appropriate OKRs as feedbacks.

Between the OKR Creation and Management module and the modelling systems: based on the requests from stakeholders. It communicates with OKR Creation and Management module for the manipulation of the OKRs;

Between the Reasoning Engine and the modelling Systems: It submits the inference requests from the stakeholders to the Reasoning Engine for performing the reasoning actions and then provides the corresponding results as feedbacks.

## 5. CASE STUDY

Based on the formalization of semantic annotations, a prototype annotation tool, SAP-KM (Semantic Annotation Plugin for Knowledge Management), has been developed to assist the demonstration of the proposed solution. This section first introduces the context of case study and then the application of formal semantic annotations in the chosen application scenario is presented

So as to show how formal semantic annotations can contribute to the semantic interoperability in a PLM environment, the life cycle of an educational combination product that is produced in a local technical production centre, named AIPL[1], has being chosen as the context of this case study. The Figure 7 shows the components of this product, which are designed to be assembled and disassembled easily. The requirements of this product are coming from the needs of reusability of the educational materials. The mechanical engineers at AIPL conceived and designed the educational combination product using the CATIA[2] Computer-Aided Design software (we name it as CATIA in the remaining paper), which generates the product technical information into a so-called Engineering Bill of Material (EBOM).

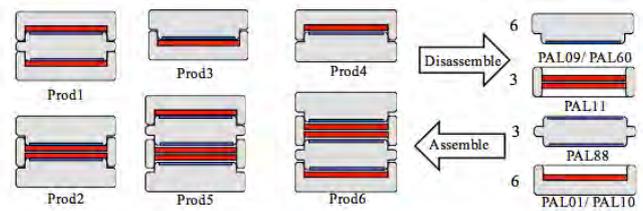

Figure 7. The Overview of the Educational Combination Product in the AIPL

However, the information in the EBOM represents the product structure from the designer's point of view, which does not contain all the information that is needed by the systems at the production stage. For this reason, a Bill of Process (BOP) needs to be combined together with EBOM. These processes are defined and modelled using the MEGA modelling environment (we name it as MEGA in the remaining paper). The Figure 8 gives a brief overview of the manufacturing processes of this product: (1) Bases turning process, which is in charge of chipping an aluminium bar into a number of designed bases. (2) Discs cutting process, which is in charge of cutting galvanized plates and magnetic plates into a number of designed discs. (3) Parts sticking process, which is in charge of using glues to stick the galvanized or magnetic discs to the corresponding bases for producing four kinds of designed parts (on the right hand side of the Figure 7). (4) Products assembling process, which is in charge of assembling different kinds of parts into six kinds of the designed products (on the left hand side of the Figure 7).

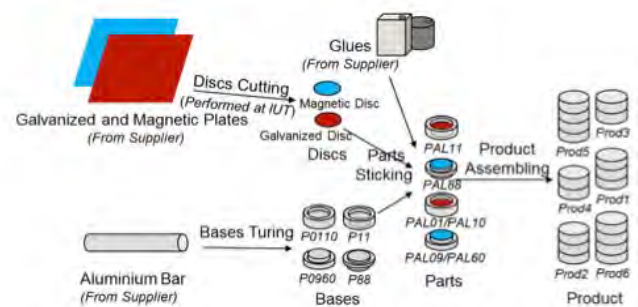

Figure 8. The Main Manufacturing Processes of the Combination Product

The EBOM and the BOP are used as basis to support the parameterization of the enterprise systems in production stage. For example, the Sage X3 ERP system[3] (we name it as Sage X3 in the remaining paper), which takes customer orders as inputs and generates work orders for supporting the purchasing of raw materials, the outsourcing of some processes and the manufacturing of the components and the related products. At the end, after some quality examination, all the qualified products are packed in boxes and dispatched to the production engineering teaching group. The information in the product life cycle is not just simply passed from one system to another in a linear unique direction. In order to differentiate the systems that are used in a selected information flow, we define three kinds of systems: the current system, which is used in a specific point of the selected information flow; the upstream system, based on the selected information flow, which is the system that is placed before the current system; the downstream system, based on the selected information flow, it is the system that is placed after the current system. So as to determine a clear-cut information flow and to show the interoperation between those systems, we choose MEGA as the current system, together with its upstream system (CATIA) and downstream system (Sage X3) as the application scenario. As shown in the Figure 9, the process model at the bottom shows all the processes that in the product lifecycle of this product. CATIA, MEGA and Sage X3 are one of the system that are used in corresponding processes respectively, which represent the TKR creation and management module in the Semantic Annotation Framework. For the other part of the framework, the Protégé is used as the OKR Creation and Management module, the Microsoft windows folder system is used as the Knowledge Cloud module, the SAP-KM is the Semantic Annotation and Processing Agency, and the Jena Reasoner employed as the Reasoning Engine Module.

---
[1] AIPL (Atelier Inter-Etablissements de Lorraine): http://www.aip-primeca.net/
[2] CATIA http://www.3ds.com/products-services/catia/
[3] Sage X3 http://www.sage.com/

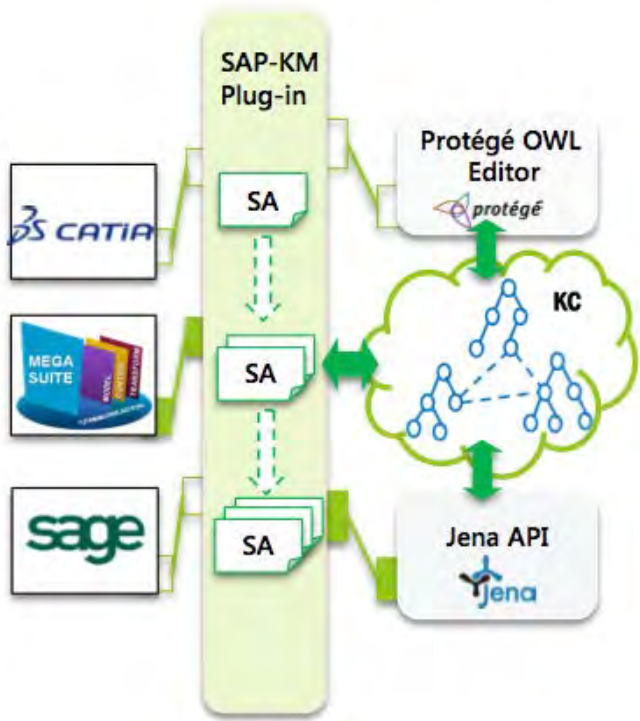

Figure 9 The Application Scenario of the Case Study

In the current version of SAP-KM, there are two developed interfaces. One is between MEGA and SAP-KM to assist the annotation on the model diagram, and the other one is between the Jena Reasoner and SAP-KM to support the inference process. In order to avoid the unnecessary repetition with several research literatures ((Attene et al., 2009), (Li et al., 2012), and (Bergamaschi et al., 2011)) which already showed the possibility of developing an annotation plug-in for product design models and data models we didn't developed the interface between CATIA/Sage X3 and SAP-KM. In the case study, we assume that the corresponding plug-ins for these two systems exists.

Based on the semantic annotation procedure, the application of the formal semantic annotation is divided into three phases: the preparation phase, the annotation phase and the reasoning phase.

Concerning the TKR part, we take into account two models: the product design model created by CATIA, which is considered as the model from upstream system that is already been annotated; and the process model created by MEGA, which is considered as the model that needs semantic enrichment.

To be more specific, the process model in the Figure 10 contains five main participants: (1) The application participant, Sage X3, which produces different kinds of orders for other participants and collects the corresponding feedbacks; (2) The warehouse, which is in charge of delivering raw materials to the work centre US (Aluminium Bars) and to the work centre CO (Galvanized Discs and Magnetic Discs). It also stores the finished component (Prod3); (3) the work centre US, which is in charge of the bases turning operation. It takes the aluminium bars as inputs and it produces two kinds of bases (P0110 and P0960); (4) the work centre CO that is in charge of the parts sticking operation. It takes the outputs of the previous operation and the raw materials from the warehouse to produce two kinds of parts (the PAL01 and the PAL60); (5) the work centre AS that is in charge of prod assembling operation. It takes the outputs from the sticking operation to produce the component (Prod3). At the end, this component is sent to the warehouse.

Concerning the OKR part, two domain level ontologies (the MSDL ontology (Ameri, 2011) and the BPMN ontology (Ghidini, 2008)) are employed. Based on them, one top level ontology (the general ontology) and two application level ontologies (the AIPL product ontology and the MEGA BPMN ontology) are created to fulfil the needs of annotation from different levels. As shown in the Figure 11, the contents in black colour are the extracted parts of these five ontologies. A number of pre-processes are carried out on these five ontologies. As shown in the Figure 11, the contents in green colour show some results of the pre-processes. To be more specific, these pre-processes are used to:

Add additional relationships: a set of additional relationships is added between the concepts in different ontologies. For example, the Object Property "hasShape" is added from the Individual "P0110" to the Individual "Cylinder".

Complete the top-level hierarchy: a set of "subClassOf" is added from the top-level classes to the Class "Thing", which are omitted by Protégé. This action is used to support the ontology loading in the Jena Reasoner.

Enrich the semantics of existing ontologies: two aspects of the semantics are formalized and added into the PLC-related ontologies (in both domain level and application level): (1) the semantics of a concept that is embedded in a general context is selected from the WordNet[4] service; (2) the semantics of a concept that is embedded in a specific context acquired from the special environment in the AIPL.

---

[4] WordNet http://wordnet.princeton.edu/

Store the ontologies: these five ontologies are stored in RDF/XML format to facilitate the ontology loading in the Jena Reasoner.

These five ontologies together the pre-process results are stored in the Knowledge Cloud. They have their own namespaces, which are different from each other. To ease the reading, the namespaces are omitted in the figure.

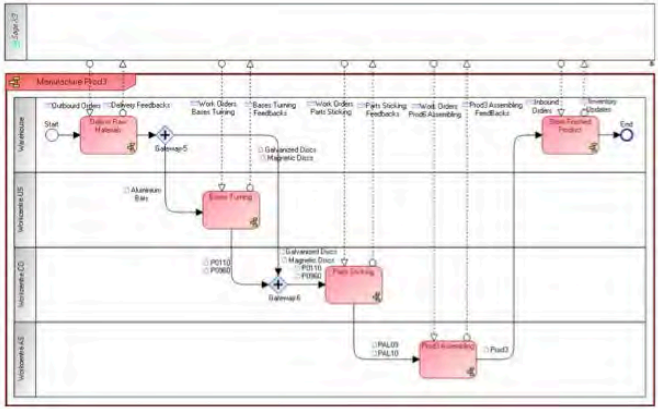

Figure 10. The Process Model from the MEGA

Based on the formal definitions of semantic annotations, a schema is designed to store the annotation results. In order to use the existing reasoning engines to assist the annotation and reasoning processes, this schema is structured as an ontology, named Semantic Annotation Schema. It uses appropriate Classes, Properties and Individuals to represent the five main elements of the Semantic Annotation and some additional Properties to assist the creation of reasoning rules. Once the preparation of OKRs, TKRs and the Semantic Annotation Schema is finished, the annotation process can be performed.

The semantic annotations that participate in the case study are divided into two parts: the received semantic annotations from the upstream system and the created semantic annotations in the current system. In order to avoid the massive details for each semantic annotation and also to ease the explanation and reading, we represent a semantic annotation in the syntax of "namespace; local name of an ontology element". The "namespace" represents the namespace of an ontology. The abbreviation namespace for General Ontology, MSDL Ontology, BPMN Ontology, AIPL Product Ontology, MEGA BPMN Ontology and Semantic Annotation Schema are respectively &GO, &MSDL, &BPMN, &AIPL, &MEGA and &SANS. The "local name" can be the local name of a class, an individual or a property.

The semantic annotations from the upstream system are imported through the SAP-KM to assist the model creation and verification in MEGA. Table 3 shows a list of model elements from the product design model with their corresponding domain semantics. In order to demonstrate the applicability of proposed solution three main operations in the manufacturing process of Prod3 ("Bases Turning", "Parts Sticking" and "Prods Assembling") together with their inputs and outputs are selected as the candidates for semantic enrichment. Concerning the explicitation of the structure semantics the internal relationships between the selected model elements are made explicit through using BPMN Ontology and MEGA BPMN Ontology. Concerning the explicitation of the domain semantics, the SAP-KM provides two possibilities: (1) to reuse the domain semantics of the imported semantic annotations through its Elements Matching function, and (2) to create new domain semantics for the selected model elements.

Table 3. The Domain Semantics of the Annotated Elements

| Model Elements | Domain Semantics | SR |
|---|---|---|
| $e_1$= 'bar' | $p_1$= &AIPL;**3mBar** &AIPL;hasLength &AIPL;3mBarLength<br>&AIPL;**3mBar** rdfs:subClassOf &AIPL;Bars<br>&AIPL;**3mBar** &AIPL;hasMaterial &MSDL;Aluminium<br>&AIPL;3mBarLength &AIPL; meters 3<br>&AIPL;Bars rdfs:subClassOf &AIPL;RawMaterial | $e_1\ sr_\subset p_1$ |
| $e_2$= '0110' | $p_2$= &AIPL;**P0110** &AIPL;hasShape &MSDL;Cylinder<br>&AIPL;**P0110** rdfs:subClassOf &AIPL;Bases<br>&AIPL;Bases rdfs:subClassOf &AIPL;SemiFiniProduct | $e_2\ sr_\subset p_2$ |
| $e_3$= '0960' | $p_3$= &AIPL;**P0960** &AIPL;hasShape &MSDL;Cylinder<br>&AIPL;**P0960** rdfs:subClassOf &AIPL;Bases<br>&AIPL;Bases rdfs:subClassOf &AIPL;SemiFiniProduct | $e_3\ sr_\subset p_3$ |
| $e_4$= 'RA' | $p_4$= &AIPL;**MDisc** &AIPL;hasShape &MSDL;Disk<br>&AIPL;**MDisc** rdfs:subClassOf &AIPL;Discs<br>&AIPL;**MDisc** &AIPL;hasMaterial &MSDL; MagneticSteel<br>&AIPL;Discs rdfs:subClassOf &AIPL;RawMaterial | $e_4\ sr_\subset p_4$ |
| $e_5$= 'RG' | $p_5$= &AIPL;**GDsic** &AIPL;hasShape &MSDL;Disk<br>&AIPL;**GDsic** rdfs:subClassOf &AIPL;Discs<br>&AIPL;**GDsic** &AIPL;hasMaterial &MSDL;GalvanizedSteel<br>&AIPL;Discs rdfs:subClassOf &AIPL;RawMaterial | $e_5\ sr_\subset p_5$ |
| $e_6$= 'Part09' | $p_6$= &AIPL;**PAL09** &AIPL;hasDiscSide &AIPL;Downward<br>&AIPL;**PAL09** rdfs:subClassOf &AIPL;Parts<br>&AIPL;Parts rdfs:subClassOf &AIPL;SemiFiniProduct | $e_6\ sr_\subset p_6$ |
| $e_7$= 'Part10' | $p_7$= &AIPL;**PAL10** &AIPL;hasDiscSide &AIPL;Upward<br>&AIPL;**PAL10** rdfs:subClassOf &AIPL;Parts<br>&AIPL;Parts rdfs:subClassOf &AIPL;SemiFiniProduct | $e_7\ sr_\subset p_7$ |
| $e_8$= 'Prod3' | $p_8$= &AIPL;**Prod3** rdfs:subClassOf &AIPL;Prods<br>&AIPL;Prods rdfs:subClassOf &AIPL;FiniProduct | $e_8\ sr_\subset p_8$ |

So as to reuse existing semantic annotations the matching between the annotated elements in the former model and the selected elements in the current model need to be implemented. After the matching process, the matched elements in the product design model have their domain semantics related to their corresponding matched elements in the process model. On the other hand the domain semantics of the selected model elements are made explicit through using General Ontology, MSDL Ontology and AIPL Product Ontology. All the semantic annotations are stored in the Semantic Annotation Schema, which are used as one of the basis for the reasoning phase.

In the case study, the reasoning phase is mainly in charge of (1) suggesting inferred semantic annotations; (2) detecting some inconsistencies between several semantic annotations of an annotated model element; and (3) identifying the possible mistakes, namely conflicts, among annotated model elements.

As shown in the Figure 11, three semantic block delimitation rules are used in the case study to make explicit the internal relationships among the annotated elements in the process model. The rule "Operation_to_DataObject" and the rule "DataObject_to_Operation" are used to create the semantic blocks that supersede the semantics between an operation and the data objects that are related to it. The rule "Operation1_to_Operation2" is used to create the semantic blocks that substitutes the semantics between two operations, which are connected by a sequence flow. These rules only show three possible situations between two appointed types of model elements. However they are enough for supporting the SBR delimitation in the case study. After the semantic block delimitation, these three object propeties are added between the corresponding ontology elements in the Semantic Annotation Schema.

```
@prefix SANS: http://www.semanticweb.org/ontologies/2013/6/SemanticAnnotations#
@prefix MEGA: http://www.semanticweb.org/ontologies/2013/6/MEGA_BPMN#
@prefix BPMN: http://dkm.fbk.eu/index.php/BPMN_Ontology#
[Operation_to_DataObject:  (?OP  rdf:type  MEGA:Operation)
                           (?DO  rdf:type  MEGA:DataObject)
                           (?SF  rdf:type  MEGA:SequenceFlow)
                           (?DO  MEGA:attachesTo  ?SF)
                           (?OP  BPMN:has_secquence_flow_source_ref_inv ?SF)
                           ->
                           (?OP  SANS:SBR_Operation_to_DataObject  ?DO)
]
[DataObject_to_Operation:  (?OP  rdf:type  MEGA:Operation)
                           (?DO  rdf:type  MEGA:DataObject)
                           (?SF  rdf:type  MEGA:SequenceFlow)
                           (?DO  MEGA:attachesTo  ?SF)
                           (?OP  bpmn: has_secquence_flow_target_ref_inv ?SF)
                           ->
                           (?OP  SANS:SBR_DataObject_to_Operation ?DO)
]
[Operation1 to Operation2:  (?OP1 rdf:type  MEGA:Operation)
                            (?SF  rdf:type  MEGA:SequenceFlow)
                            (?OP2 rdf:type  MEGA: Operation)
                            (?SF  BPMN:SourceRef  ?OP1)
                            (?SF  BPMN:TargetRef  ?OP2)
                            ->
                            (?OP1  SANS:SBR_Operation1_to_Operation2 ?OP2)
]
```

Figure 11 Three Rules to define a SBR for making explicit the Relations

The property association process is performed between the properties, which are made explicit in the semantic block delimitation process, and the properties in the PLC-related ontologies. Based on the inferred semantic annotations suggestion algorithm corresponding inferred semantic annotations are suggested. After the suggestion, the comparison of the similarity between two domain semantics of a common annotated model element can be performed. SAP-KM queries all the individuals that have both initial and inferred semantic annotations in the Class "&SANS;E" and it generates all the possible comparison pairs between an initial one to an inferred one. The similarity comparison results and the inconsistency detection rules are used as inputs of the reasoning engine to produce the inconsistency detection results. The inconsistency detection results are used as inputs of the model conflict identification rules and algorithms. The possible conflicts between annotated model elements are used to draw attention of modellers for examining the correctness of two annotated elements in the process model. Ideally, the model content conflicts identification results are supposed to contain the reason why two model elements have conflicts and how to solve these possible mistakes. However, these kinds of suggestions highly rely on the power of the reasoning engine. The current reasoning engines are only able to deal with some simple reasoning, such as classification (class subsumption and individual memberships) and class consistency (whether a class can have individuals or not), but they cannot deal with sub-ontologies.

The process model and the created semantic annotations are sent to Sage X3 to assist the parameterization. Let us take the table of "process planning" in Sage X3 as an example. The "process", "work centre", "preparation time" and "execution time" are the four of its main elements in the parameterization. Concerning the operation "Bases Turning" ($e_9$), "Parts Sticking"($e_{15}$) and "Product Assembling" ($e_{21}$), the corresponding that are need by Sage X3 are contained in their the semantic annotations $p_{10}$, $p_{12}$ and $p_{14}$ respectively. Once the semantic annotations are created in the Sage X3 data model, the corresponding elements matching in the Sage X3 plug-in is able to assist the stakeholder to fill the right data into the right fields of the "process planning" table. This case study shows how the formal semantic annotations are contributing in: (1) acquiring the initial semantics that the stakeholders, who manipulate the upstream system, wanted to express; (2) verifying, semi-automatically, the semantic consistency between the contents in the received models and in the developing models; and (3) guaranteeing the correctness of the embedded semantics in the under development models for the stakeholders, who manipulate the downstream system.

## 6. CONCLUSION

The proposed solution in this paper provides some fundamental contributions, which are remarked as follows: (1) A semantic annotation meta-model that unambiguously describes the major components of a semantic annotation and their interrelations; (2) The definition of the Semantic blocks for the semantics description and substitution; (3) Formal definitions of the semantic annotation; (4) Three reasoning mechanisms that show and validate the usages of semantic enrichments; (5) A semantic annotation procedure; (6) A semantic annotation framework. Concerning the hypothesis one, although the research focus of this work is not in capturing, representing and formalizing knowledge into ontologies, the richness of OKRs in the Knowledge Cloud influences the precision of the semantic annotations. Concerning the hypothesis two, the interconnections among ontologies are the fundamental of using multiple ontologies together to perform semantic enrichment and perform inference on semantic annotations. Reasoning engines are not able to perform reasoning on concepts coming from different ontologies which have not relationships (directly or indirectly) between each other. Concerning the hypothesis three, the semantic similarities between two domain semantics are used as the basis to support the inconsistency detection. The more precise semantic similarities are the more precise results can be produced. From the practical point of view, the prototype implementation and validation shows the possibility of using the formalization of semantic annotations for system interoperability in a PLM environment. Furthermore, in the context of a PLM environment, three interesting directions can also be considered as future works. (1) To enable the traceability of requirements. With the assistance of semantics annotation, it is possible to trace the validation of each requirement in every stage of the product lifecycle, from the initial design until the final deposit of. (2) To make explicit the relationships between the TKRs. (3) To address the versioning of models. The issue about the versioning of models in a PLC is difficult to be avoided. Semantically enriching models gives the possibility to ensure that the modified model contents is not semantically in conflict with existing ones.

In a nutshell, the purpose of this work is to deal with the issue of semantic interoperability. Despite some limitations, as discussed in this section, we are convinced that the proposed formalization of semantic annotations is able to support and guarantee the models interoperability in a PLM environment.

REFERENCES

Ackoff, R.L.: From Data to Wisdom. J. Applies Syst. Anal. 16, 3–9 (1989).

Ameri, F., Dutta, D.: Product lifecycle management: closing the knowledge loops. Comput.-Aided Des. Appl. 2, 577–590 (2005).

Ameri, F., McArthur, C., Asiabanpour, B., Hayasi, M.: A web-based framework for semantic supplier discovery for discrete part manufacturing. SME/NAMRC. 39, (2011).

Attene, M., Robbiano, F., Spagnuolo, M., Falcidieno, B.: Characterization of 3D shape parts for semantic annotation. Comput.-Aided Des. 41, 756–763 (2009).

Bergamaschi, S., Beneventano, D., Corni, A., Kazazi, E., Orsini, M., Po, L., Sorrentino, S.: The Open Source release of the MOMIS Data Integration System. Proc. of the Nineteenth Italian Symposium on Advanced Database Systems, SEBD. pp. 26–29 (2011).

Boudjlida, N., Dong, C., Baïna, S.: A practical experiment on semantic enrichment of enterprise models in a homogeneous environment. (2006).

Boudjlida, N., Panetto, H.: Annotation of enterprise models for interoperability purposes. In: IEEE (ed.) IEEE International Workshop on Advanced Information Systems for Enterprises, IWAISE'2008. pp. 11–17. , Constantine, Algeria (2008).

Di Francescomarino, C.: Semantic annotation of business process models, (2011).

Doan, A., Madhavan, J., Dhamankar, R., Domingos, P., Halevy, A.: Learning to match ontologies on the Semantic Web. VLDB J. 12, 303–319 (2003).

Elgueder, J., Cochennec, F., Roucoules, L., Rouhaud, E.: Product–process interface for manufacturing data management as a support for DFM and virtual manufacturing. Int. J. Interact. Des. Manuf. IJIDeM. 4, 251–258 (2010).

Euzenat, J.: Towards a principled approach to semantic interoperability. Presented at the IJCAI 2001, workshop on ontology and information sharing , Seattle, USA August (2001).

Gašević, D., Devedžić, V.: Petri net ontology. Knowl.-Based Syst. 19, 220–234 (2006).

Ghidini, C., Rospocher, M., Serafini, L.: A formalisation of BPMN in description logics. Technical report TR 2008-06-004, FBK-irst (2008).

Gruber, T.R.: A translation approach to portable ontology specifications. Knowl. Acquis. 5, 199–220 (1993).

IEEE Standard Computer Dictionary. A Compilation of IEEE Standard Computer Glossaries. IEEE Std 610. 1– (1991).

Li, C.: Ontology-Driven Semantic Annotations for Multiple Engineering Viewpoints in Computer Aided Design, (2012).

Liao, Y., Lezoche, M., Rocha Loures, E., Panetto, H., Boudjlida, N.: Semantic Enrichment of Models to Assist Knowledge Management in a PLM environment. 21 st International Conference on Cooperative Information Systems (CoopIS 2013). , Graz, Austria (2013)

Maedche, A., Staab, S.: Measuring Similarity between Ontologies. Presented at the 13th International Conference on Knowledge Engineering and Knowledge Management: Ontologies and the Semantic Web , Sigüenza, Spain (2002).

Miller, J., Mukerji, J.: MDA Guide. Object Manag. Group. 234, 51 (2003).

Nonaka, I.: A dynamic theory of organizational knowledge creation. Organ. Sci. 5, 14–37 (1994).

Patil, A.A., Oundhakar, S.A., Sheth, A.P., Verma, K.: Meteor-s web service annotation framework. Proceedings of the 13th international conference on World Wide Web. pp. 553–562 (2004).

Pokraev, S., Quartel, D., Steen, M.W.A., Reichert, M.: Semantic Service Modeling: Enabling System Interoperability. In: Doumeingts, P.G., Müller, P.J., Morel, P.G., and Vallespir, P.B. (eds.) Enterprise Interoperability. pp. 221–230. Springer London (2007).

Polanyi, M.: The tacit dimension. Peter Smith Gloucester, MA (1966).

Rink, D.R., Swan, J.E.: Product life cycle research: A literature review. J. Bus. Res. 7, 219–242 (1979).

Stumme, G., Maedche, A.: Ontology merging for federated ontologies on the semantic web. Presented at the FMII-2001, International Workshop for Foundations of Models for Information Integration , Viterbo, Italy September (2001).

Uren, V., Cimiano, P., Iria, J., Handschuh, S., Vargas-Vera, M., Motta, E., Ciravegna, F.: Semantic annotation for knowledge management: Requirements and a survey of the state of the art. Web Semant. Sci. Serv. Agents World Wide Web. 4, 14–28 (2006).

Yahia, E., Lezoche, M., Aubry, A., Panetto, H.: Semantics enactment for interoperability assessment in Enterprise Information Systems. Annu. Rev. Control. 36, 101–117 (2012).

W3C: SAWSDL, http://www.w3.org/TR/2007/REC-sawsdl-20070828/.

Zdravković, M., Panetto, H., Trajanović, M., Aubry, A.: An approach for formalising the supply chain operations. Enterp. Inf. Syst. 5, 401–421 (2011).